# A Large Area Timing RPC

A. Blanco[1,2], R. Ferreira-Marques[1,3], Ch. Finck[4], P. Fonte[1,5,*], A. Gobbi[4], A. Policarpo[1,3], M. Rozas[2].

1-LIP, Coimbra, Portugal.
2-GENP, Dept. Fisica de Particulas, Univ. de Santiago de Compostela, Espanha
3-Departamento de Física da Universidade de Coimbra, Portugal.
4-GSI, Darmstadt, Germany.
5-ISEC, Coimbra, Portugal.

## Abstract

A large area Resistive Plate Chamber (RPC) with a total active surface of $160\times10$ cm$^2$ was built and tested. The surface was segmented in two 5 cm wide strips readout on both ends with custom, very high frequency, front end electronics.

A timing resolution between 50 and 75 ps $\sigma$ with an efficiency for Minimum Ionizing Particles (MIPs) larger than 95% was attained over the whole active area, in addition with a position resolution along the strips of 1.2 cm . Despite the large active area per electronic channel, the observed timing resolution is remarkably close to the one previously obtained (50 ps $\sigma$) with much smaller chambers of about 10 cm$^2$ area.

These results open perspectives of extending the application of timing RPCs to large area arrays exposed to moderate particle multiplicities, where the low cost, good time resolution, insensitivity to the magnetic field and compact mechanics may be attractive when compared with the standard scintillator-based Time-of-Flight (TOF) technology.

[*] Corresponding author: Paulo Fonte, LIP - Coimbra, Departamento de Física da Universidade de Coimbra, 3004-516 Coimbra, PORTUGAL.
tel: (+351) 239 833 465, fax: (+351) 239 822 358, email: fonte@lipc.fis.uc.pt

# 1 Introduction

The development of timing Resistive Plate Chambers (RPCs) [1] opened the possibility to build large-granularity high-resolution TOF systems at a quite reduced cost per channel when compared with the standard scintillator-based technology.

Previous work has shown a timing resolution better than 50 ps $\sigma$ at 99% efficiency in single four-gap chambers [2] and an average timing resolution of 88 ps $\sigma$ at and average efficiency of 97% in a 32 channel system [3]. It has also been shown that each amplifying gap of 0.3 mm thickness has a detection efficiency close to 75% and that the avalanche develops under the influence of a strong space charge effect [4]. A Monte-Carlo model of the avalanche development reproduced well the observed data, confirming the dominant role of space charge effects in these detectors [5].

Although timing RPCs have so far been built with relatively small active areas per electronic readout channel (on the order of 10 $cm^2$), compatible with the high-multiplicity requirements of High Energy Heavy Ion Physics, there is a number of possible applications in lower multiplicity environments ([6], [7]) for more coarsely segmented counters.

Having in view such applications, we describe in this paper the structure and performance of a large counter, with an active area of $160 \times 10$ $cm^2$, readout by only 2 or 4 electronic channels.

# 2 Detector description

The detector was built from 3 mm thick float-glass plates with an area of $160 \times 12$ $cm^2$ and a measured bulk resistivity of $2 \times 10^{12}$ $\Omega$ cm. The stack of plates, with attached copper foil[1] electrode strips (see Figure 1), was mounted on a supporting 1 cm thick acrylic plate and a controlled pressure was applied to the stack by means of regularly spaced spring-loaded plastic bars. Four gas gaps were defined by glass fibres of 0.3 mm diameter placed between the glass plates, beneath the pressing bars.

There were six individual electrode strips, with dimensions of $160 \times 5$ $cm^2$, connected in two independent multilayer groups with a 1 mm wide interstrip distance. The arrangement defined an active area of $160 \times 10$ $cm^2$, leaving uncovered a 1 cm wide region on the outer rim of the glass plates.

The ends of each group were connected via a short 50 $\Omega$ coaxial cable (Figure 1) to preamplifiers placed inside the gas volume, whose signal was fed through gas tight connectors to external amplifier-discriminator boards. The front-end chain was custom-made from commercially available analogue and digital integrated circuits, yielding measured timing and charge resolutions of respectively 10 ps $\sigma$ and 3.2 fC $\sigma$ [8]. For stability reasons the discriminators had a built-in dead time of 1 µs after each detected pulse that will contribute to the overall counter inefficiency.

High voltage, around 6 kV, was applied to the outer strips via 10 M$\Omega$ resistors and the signal travelling in these strips was fed to the shielding of the signal cables via 2.2 nF high voltage capacitors (see Figure 1). The central wire of the signal cables was connected directly to the central strips and to the preamplifiers inputs. To avoid the use of floating electrodes, the glass plates placed in the middle of each (upper and lower) half of the detector had thin copper electrodes glued along their lateral edges, kept by a resistive voltage divider at half of the voltage applied to the outer strips.

[1] 3M conductive-adhesive copper tape.

The detector, high voltage distribution network and preamplifiers were placed inside a gas-tight aluminium enclosure that was kept under a continuous flow of a non-flammable gas mixture consisting of 85% $C_2H_2F_4$[†] + 10% $SF_6$ + 5% iso-$C_4H_{10}$ [9], at a flow rate close to 100 $cm^3$/min.

# 3 Test Setup and Data Acquisition

The tests were made at the CERN PS using a secondary beam (T11) of 3.5 GeV particles, mainly negative pions, in August 2000. The spills were 0.3 s long and spaced by a few seconds. Most tests were done with a defocused beam that illuminated the detector over a region of a few hundred $cm^2$.

A pair of plastic scintillation counters (Bicron BC420) measuring $8\times3\times2$ $cm^3$ and viewed on each $3\times2$ $cm^2$ face by a fast photomultiplier (Hamamatsu H6533) provided the reference time information. Both counters had a timing resolution close to 35 ps $\sigma$ and defined a coincidence (trigger) area of $2\times2$ $cm^2$, being placed upstream from the RPC.

The data acquisition system, based on the CAMAC standard, was triggered by the coincidence of both timing scintillation counters, being additionally required that no signal was present (veto) in a fourth wide scintillation counter that surrounded the coincidence-selected beam.

After a valid trigger the four timing signals from the reference scintillation counters and the four timing signals from the RPC were digitised by a LeCroy 2229 TDC. A LeCroy 2249w ADC, operated with a gate of 300 ns, digitised the corresponding charge information. The TDC was calibrated using an Ortec 240 time calibrator and the ADC was calibrated by injecting in the preamplifiers a known amount of charge via the test inputs.

# 4 Data Analysis

Prior to any analysis the TDC digitisation error was taken into account by adding to each data value a random number distributed in the interval [-0.5, 0.5] and the events were selected by external and internal cuts.

After cuts the events were attributed to the strip showing the largest charge signal and the data from each strip was analysed separately.

## 4.1 External Cuts

To clear the data from beam-related artefacts, like multi-particle events or scattered particles, several cuts were made based on the information collected from the timing scintillators. Events selected for further analysis had to comply with the following requirements:

† Commercially known as R134a.

- the difference between the TOF information from both timing scintillators (time average between both ends of a counter) should agree within $\pm 2\sigma$ of the mean value;
- the position information from each timing scintillator (time difference between both ends of a counter) should be within $\pm 2\sigma$ of the respective mean value[2];
- the charge measured on each timing scintillator should lie between the 5% and the 95% percentiles of the respective (strongly non-gaussian) distribution.

All these distributions showed wide tails beyond the specified cuts, generating corresponding tails in the time response of the RPC (see section 4.5).

Typically about 50% of the events were accepted after external cuts.

### 4.2 Internal Cuts

The time difference between both strip ends (that should be independent of the details of the avalanche development process and depend only on its position) shows large bilateral tails that were cut at $\pm 2\sigma$ of the mean value[3]. This cut had a negligible effect on the time resolution of the counter but it strongly reduced the amount of timing tails (see section 4.5) close to the strip ends. Tighter cuts have little further influence and presumably the remaining timing tails are mostly due to the physical process related to the avalanche onset and development (depending, for instance, on the applied voltage - see section 5.4).

It should be noted that when the quality of the trigger was enhanced by requiring an additional coincidence with a small ($0.3\times 0.6$ cm$^2$) scintillator placed downstream from the RPC the amount of tails became negligible (see Figure 8) and the cut mentioned above had little effect on the results, suggesting that the need for such cut arises mainly due to beam quality limitations.

For accurate correction of the measured time as a function of charge and to decrease the amount of timing tails seen by the RPC without significant efficiency loss, a 1% cut was made in the lower part of the RPC charge distribution (Figure 2). (The RPC charge was defined as the sum of the charges sensed on both ends of each strip.)

### 4.3 Detection efficiency

The detection efficiency was determined for each strip after the external cuts and before the internal cuts. The impact of the (optional) internal cuts should be subtracted from the efficiency values given in all figures.

Two different definitions of detection efficiency were formed by the ratio of the following quantities to the total number of events:

- the number of events yielding an amount of charge in the RPC larger than the upper limit of the ADC pedestal distribution (charge efficiency);
- the number of events for which a valid time was measured in the RPC (time efficiency).

---

[2] The position spread was mainly related with the width of the $2\times 2$ cm$^2$ coincidence-selected trigger region.
[3] Note that the same type of cut was applied to the timing scintillator data.

In principle both definitions should yield similar results except if a considerable amount of inter-strip crosstalk will be present. In this case the crosstalk signal, not being galvanically coupled, will not contribute to any net charge but may induce a voltage level above the discriminating threshold, generating a valid-time event.

### 4.4 Position accuracy

A position-dependent timing information can be formed for each event by the time difference between both strip ends. This information was calibrated with respect to a known displacement of the RPC and it was charge-corrected in a manner similar to the TOF information (section 4.5).

### 4.5 Timing accuracy

In principle a position-independent timing (TOF) information can be formed for each event by averaging the time measured on both strip ends. However, it was found that close to the extremes of the counter the TOF information was correlated with the position information and a linear correction could be applied to the former as a function of the later. Data will be presented with and without this position correction.

The TOF information also strongly correlates (see Figure 3) with the measured signal charge and a correction was made along the lines described in [2]. The method automatically determines and corrects the contribution of the reference counters time jitter.

The resulting time distribution was not purely gaussian, showing a bilateral excess of events (timing tails) exemplified in Figure 4. We opted to characterize separately the main timing resolution figure ($\sigma$) and the amount of timing tails, since the later can be important or not, depending on the application.

The main resolution figure was determined by a gaussian fit to the corrected time distribution (with 5 ps bins) within $\pm 1.5\ \sigma$ of the mean value and the timing tails were characterized as the fraction of events whose distance to the mean value exceeds 300 ps. For a purely gaussian distribution with $\sigma \leq 100$ ps this fraction should be smaller than 0.3%.

For each run the time vs. log(charge) correlation curve, which is almost linear, was characterised by the average time, charge and slope (see Figure 3). This information was used to assess the stability of the time-charge correlation and the need for separate correction curves at different positions along the counter.

## 5 Results and Discussion

In the following discussion we will refer to the position of the centre of the trigger region along and across the strips as, respectively, the X and Y coordinates, defining the origin of the coordinate system in the geometrical centre of the counter.

### 5.1 Detection Efficiency and Crosstalk

Charge and time efficiency curves are shown in Figure 5 a) as a function of Y (for X=0). The fraction of events inducing a measurable amount of charge in both strips is also shown, probably corresponding to avalanches occurring close to the inter-strip region.

When the trigger region was fully contained within a single strip the charge and time efficiencies closely match for that strip, while the opposing strip shows a very reduced charge efficiency and a considerable (from 80% to 90%) time efficiency. This large crosstalk level (see section 4.3), actually to be expected on such long strips, did not significatively affect the timing characteristics of the device, but would affect its multi-hit capability.

In Figure 6 a) the time efficiency is shown as a function of X (for Y=±3 cm, corresponding essentially to the centre of each strip). The measurements were generally taken in steps of 7.5 cm except for a region of strip A, between 0 and 20 cm, that was scanned in steps of 1.5 cm to assure that at least one measurement contained a spacer.

The values range between 95% and 98%, being slightly larger for the strip A. This small difference can be attributed to slight differences in the gain of the front-end electronics chain. However it should be noted that a smaller chamber of similar construction has shown a time efficiency above 99% [2]. The slightly reduced efficiency found in the present counter can be attributed to a poorer trigger quality, evidenced by the tails visible in the scintillators time and position information (see section 4.1) and to a much larger sensitive area that collects a larger event rate from the wide beam (see the discussion about the discriminators dead-time in section 3).

The combination of both strip signals into a single amplifier for each end of the counter, doubling the active area per amplifier, caused absolutely no degradation in the detection efficiency. Also no influence from the spacers could be found in the fine-step scan.

## 5.2 Timing Accuracy

### *5.2.1 Timing resolution*

The timing resolution is shown as a function of Y (for X=0) in Figure 5 b). It ranges from 58 to 76 ps $\sigma$ across the counter, including the outer edges and the inter-strip region. Since in a real application there would be no possibility to determine the avalanche position along Y, in the same figure we present also (horizontal lines) the resolution figure obtained by analysing for each strip a data set containing an equal number of events from each position, yielding 67 and 76 ps $\sigma$ for the strips A and B, respectively.

In Figure 6 b) we show the timing resolution as a function of X (for Y=±3 cm). Most data points range from 50 to 70 ps $\sigma$, except for two regions, around -20 cm and -70 cm, where the resolution was degraded to, respectively, 80 and 90 ps $\sigma$. This degradation most probably has a local mechanical origin, since it is not symmetrical with the counter geometry and not identical for both strips.

The combination of both strip signals into a single amplifier for each end of the detector, doubling the active area per amplifier, caused absolutely no degradation in the time resolution of the device. Also no clear influence from the spacers could be found in the fine-step scan.

### *5.2.2 Timing tails*

The magnitude of the timing tails is shown as a function of Y (for X=0) in Figure 5 c). The tails do not exceed 2% of the total number of events, being smaller than 1% in the centre of the strips. The effect is larger in the outer edges than in the inner strip edges, possibly because no attempt was made to sharply reduce the electric field at the strip edges (the glass plates extend up to 1 cm beyond the copper strips). Avalanches occurring in this space will induce currents both in the strips and in the enclosing gas box, creating a position-dependent induced charge fluctuation that may cause timing errors. A similar phenomenon can be perceived in the space between the strips, where the induced charge was shared among the strips in a position-dependent manner (see also Figure 5 a)).

In Figure 5 c) we present also (horizontal lines) the values obtained by analysing for each strip a data set containing an equal number of events from each position, yielding tails of 1% for both strips.

In Figure 6 c) we show the amount of timing tails as a function of X (for Y=±3 cm). Strip B shows tails generally below 1.0 %, raising up to 1.5 % close to the counter extremities. Strip A shows larger tails, up to 2% over the whole counter. A possible reason for this difference, that doesn't appear in the Y-scan (Figure 5 c)), could be a momentary beam quality fluctuation.

The combination of both strip signals into a single amplifier for each side of the detector, doubling the active area per amplifier, caused absolutely no degradation in the amount of tails and no clear influence from the spacers could be found in the fine-step scan.

#### *5.2.3 Variations of the measured time and charge along the counter*

Due to mechanical or electrical inhomogeneities there will be position dependent charge and time variations along the counter. This effect should be corrected by calibration using the counter's position resolution (discussed in section 4.4), being however interesting to determine how finely segmented this calibration should be.

In Figure 7 we show the variations of time (a), charge (b) and of the slope of the time vs. log(charge) correlation curve (c), represented in Figure 3, as a function of X (for Y=±3 cm). Variations of the average time by about 400 ps are apparent along with large changes of the time vs. log(charge) correlation slope, while the average charge remains relatively stable. It should be noted that the large slope variations visible in the left-hand side of Figure 7 c) correlate well with the degradation of the time resolution visible in the same region of Figure 6 b).

To evaluate directly the influence of these position dependencies on the time resolution of the counter as a function of the segmentation of the calibration procedure along X, we have combined an equal number of events from adjacent positions along the strip B and analysed jointly the resulting data set. The results are shown in Figure 7 d): events from the right-hand side of the counter could be jointly analysed without much degradation arising from position dependent effects, while the left hand side was severely affected by such effects, calling for a finer segmentation of the calibration procedure. Such features have probably a local, mechanical, origin, since all other variables are equal along the counter.

### 5.3 Position Resolution

In Figure 8 a) the time difference between both strip ends is plotted as a function of X (for Y=±3 cm) showing an accurately linear dependency with a slope of 70.9 ps/cm, which corresponds to a signal propagation velocity of 14.1 cm/ns. In Figure 8 b) the time difference distribution is plotted for two trigger positions 5.0 cm apart, yielding a position accuracy of 12 mm σ. For this measurement the width of the trigger region in the X direction was reduced to 3 mm via an extra coincidence scintillator.

### 5.4 Behaviour as a function of the applied voltage

It is interesting to study how some of the quantities mentioned above change as a function of the applied voltage, as illustrated in Figure 9.

Figure 9 a) shows the evolution of the time resolution and efficiency, the later showing a plateau of 98% above 6.1 kV while the former shows a broad minimum at 52 ps σ close to 6.1 kV. This voltage has been chosen as the optimum operating point and most of the data presented in the previous sections has been taken at this setting.

Figure 9 b) shows the evolution of the average fast charge and of the amount of timing tails. The behaviour of charge as a function of voltage was already discussed at length in [4] and we will not further elaborate on this subject here. The amount of timing tails decreases with increasing voltage and reaches a plateau of 1% above 5.9 kV.

Figure 9 c) shows the variation of the absolute measured time and of the time vs. log(charge) correlation slope. As expected, the measured time shows a negative variation, compatible with a larger value of the avalanches exponential growth parameter; further details on this subject can be found in ([4], [5], [8]). The time vs. log(charge) correlation slope becomes less steep with increasing voltage, reaching a plateau above 6.3 kV.

### 5.5 Behaviour as a function of the counting rate

Being the counting rate capability an important characteristic of RPCs, several quantities of interest were studied as a function of the counting rate per unit area, as show in Figure 10. The vertical arrow indicates the standard operating point (140 Hz/cm$^2$) at which most of the measurements presented above were taken.

Figure 10 a) shows the variation of the time resolution and efficiency. Both quantities are constant below 140 Hz/cm$^2$ and degraded at larger counting rates. Operation up to 500 Hz/cm$^2$ may be possible if a slight degradation of the counter performance is accepted (comparable with the performance variations observed as a function of position).

Figure 10 b) shows the evolution of the average fast charge and of the amount of timing tails. The average fast charge shows a continuous decrease with increasing counting rate, suggesting a counting-rate induced reduction of the average electric field in the amplifying gap, while the timing tails remain quite small (less than 4 %) up to 1 kHz/cm$^2$.

Figure 10 c) shows the variation of the absolute measured time and of the time vs. log(charge) correlation slope. The average measured time shows a positive variation, compatible the observed reduction of the gas gain, which is nevertheless quite small (±10 ps) around the standard operating point, while the slope remains essentially unchanged around this point. It should be stressed that any variations of these quantities can be taken into account by appropriate calibration.

The general behaviour of the detector characteristics as a function of the counting rate indicates that there is no degradation of the performance up 140 Hz/cm$^2$ and that a counting rate of 500 Hz/cm$^2$ could be handled if a slight performance degradation of will be accepted.

## 6 Conclusions

We built and tested a large area timing RPC, with an active surface of 160×10 cm$^2$, to be applied in medium (e.g. [6]) or low multiplicity (e.g. [7]) TOF counters. The active area was segmented in two readout strips, each measuring 160×5 cm$^2$, sensed in both ends by identical custom-made, very high-frequency, front-end electronic channels [8].

A timing resolution between 50 and 90 ps $\sigma$ with an efficiency between 95% and 98% for MIPs was attained over the whole active area. The performance could be improved close to the strip ends by correcting the measured time as a function of the measured avalanche position, thus improving the time resolution range to lie between 50 and 75 ps $\sigma$.

The combination of both strip signals into a single amplifier for each end of the detector, doubling the active area per amplifier, caused absolutely no degradation in the efficiency or in the time resolution of the device. Also no clear influence from the spacers could be found in a fine-step scan.

The avalanche position along the counter could be determined from the time difference between both strip ends, yielding a position resolution of 1.2 cm $\sigma$ along the strips with very good linearity.

Timing tails, defined as the fraction of events whose absolute time deviation from the average was larger than 300 ps, were smaller than 2%, occurring the larger values in the outer edges of the detector and close to the strip ends.

The general behaviour of the detector characteristics as a function of the counting rate indicates that there is no degradation of the performance up 140 Hz/cm$^2$ and that a counting rate of 500 Hz/cm$^2$ could be handled if a slight degradation is accepted.

Probably due to structural inhomogeneities there were considerable variations of the average measured time along the counter, requiring a calibration procedure segmented every few tens of centimetres. In an experimental array this segmentation would be achieved with the help of the counter's position resolution.

The large inter-strip crosstalk level observed (80%) does not seem to influence the time resolution of the counter, affecting only its multi-hit capability. It should be pondered whether for a given application it is not preferable to base the detector on multiple layers of single-strip chambers, reaching full geometrical coverage and avoiding any crosstalk. A multilayer configuration, providing multiple measurements for each particle, would also have the advantages of being self-calibrating and allowing an improved rejection of timing tails

## 7 Acknowledgements

We are grateful to Paolo Martinengo and Piotr Szimanski of the ALICE test beam support team for their efficient and friendly cooperation; to Juan Garzon from University of Santiago de Compostela for his interest and support; to José Pinhão, Américo Pereira and Fernando Ribeiro from our technical staff for their competent collaboration.

This work was done in the framework of the FCT project CERN/P/FIS/15198/1999.

# 8 References

[1] P.Fonte, A. Smirnitski and M.C.S. Williams, “A New High-Resolution Time-of-Flight Technology”, *Nucl. Instr. and Meth. in Phys. Res. A*, 443 (2000) 201.

[2] P. Fonte, R. Ferreira Marques, J. Pinhão, N. Carolino and A. Policarpo “High-Resolution RPCs for Large TOF Systems“, *Nucl. Instr. and Meth. in Phys. Res. A*, 449 (2000) 295.

[3] A. Akindinov et al.,”A Four-Gap Glass-RPC Time of Flight Array with 90 ps Time Resolution”, ALICE note ALICE-PUB-99-34, preprint CERN-EP-99-166.

[4] P.Fonte and V.Peskov. “High-Resolution TOF with RPCs”, presented at the “PSD99 - 5th International Conference on Position-Sensitive Detectors”, 13-17 th September 1999, University College, London, preprint LIP/00-04 http://xxx.lanl.gov/abs/physics/0103057.

[5] P.Fonte, “High-Resolution Timing of MIPs with RPCs – a Model”, presented at the “RPC99 - 5th International Workshop on Resistive Plate Chambers”, 28 - 29th October 1999, Bari, Italy, *Nucl. Instr. and Meth. in Phys. Res. A*, 456 (2000) 6.

[6] FOPI-Collaboration, "Upgrading the FOPI Detector System", *GSI-Scientific Report*, (1998), pp. 177.

[7] The HARP collaboration ( PS214 ), "The Hadron Production Experiment at the PS", CERN-SPSC/99-35, SPSC/P315, 15 November, 1999.

[8] A. Blanco, N. Carolino, P. Fonte, A. Gobbi, “A Simplified and Accurate Front-End Electronics Chain for Timing RPCs”, presented at the “LEB 2000-6th Workshop on Electronics for LHC Experiments”, 11-15 September 2000, Cracow, Poland, published in the conference proceedings CERN 2000-010 CERN/LHCC/2000-041.

A. Blanco, N. Carolino, P. Fonte, A. Gobbi, “A New Front-End Electronics Chain for Timing RPCs”, presented at the “2000 IEEE Nuclear Science Symposium and Medical Imaging Conference”, 15-20 October 2000, Lyon, France, accepted for publication in *IEEE Trans. Nucl. Sci.*

[9] P. Camarri et al, “Streamer suppression with $SF_6$ in RPCs operated in avalanche mode”, *Nucl. Instr. and Meth. in Phys. Res. A* 414 (1998) 317.

## 9 Figure Captions

Figure 1: Pictures and schematic drawings of the detector.

Figure 2: Typical fast charge[4] distribution in logarithmic and linear (inset) scales.

Figure 3: Typical time vs. log(charge) correlation plot, showing the calculated time-charge correction curve (thin line) and the average slope (thick line).

Figure 4: Typical time distribution (from strip B at X=-30 cm) after charge correction in logarithmic and linear (inset) scales. The thick line corresponds to a gaussian curve fitted within ±1.5 σ to determine the main resolution figure (after correction for the contribution of the start counters – see section 4.5 – the timing resolution for this example is $s_t$=53 ps). The dashed line corresponds to the extension of the fitted gaussian to ±3.5 σ. Timing tails were defined as the fraction of events whose absolute time deviation from the average was larger than 300 ps and amount in this example to 0.4 %.

Figure 5: Several quantities of interest as a function of the position of the centre of the trigger region across the strips. a) Charge and time efficiency. The lower curve corresponds to the fraction of events that show a measurable amount of charge in both strips. The superimposed dashed lines indicate the position of the copper strips and the outer dotted line the edge of the glass plates. b) Timing resolution with separate analysis in each position and when all events for each strip are analysed simultaneously. The solid triangles correspond to data taken with both strips connected together. c) Same as b), for the timing tails.

Figure 6: Several quantities of interest as a function of the position of the centre of the trigger region along the strips. a) Time efficiency, better than 95%; b) Time resolution with and without position correction, ranging from 50 to 90 ps σ (50 to 75 ps σ with position-dependent time correction). c) Timing tails with and without position correction, smaller than 2%. In all figures the solid triangles correspond to data taken with both strips connected together.

[4] Electronic component of the signal.

Figure 7: Several quantities related with the time-charge correlation curve (see Figure 3) are plotted as a function of the position of the trigger region along the strips[5]. a) Variations of the average measured time, covering a range of 400 ps. b) Average measured charge, which shows little variation along the counter. c) The average slope shows considerable variations along the counter, particularly in the left hand side, where a poorer time resolution is also visible (Figure 6). d) Joint analysis of a data set containing an equal number events from each of the positions indicated by the extent of the horizontal lines: the right hand side is only weakly affected by position dependent effects, while the left hand side would require a more finely segmented calibration procedure.

Figure 8: a) Time difference between both strip ends as a function of the position of the trigger region along the strips. There is an accurately linear dependency (evidenced by the small residues shown in b)), with a slope of 70.9 ps/cm, corresponding to a signal propagation velocity of 14.1 cm/ns. c) The width of the trigger region was reduced to 3 mm in the X direction and the spread of the time difference was compared with a measured displacement of 5.0 cm, yielding a position accuracy of 12 mm $\sigma$.

Figure 9: Several quantities of interest plotted as a function of the applied voltage. a) Time resolution and efficiency. b) Average amount of fast charge and the amount of timing tails. c) The variation of the absolute value of the measured time and the slope of the time vs. log(charge) correlation curve.

Figure 10: Several quantities of interest plotted as a function of the counting rate density. a) Time resolution and efficiency. b) Average fast charge and amount of timing tails. c) The variation of the absolute value of the measured time and the slope of the time vs. log(charge) correlation curve. The vertical arrow indicates the counting rate at which most of the measurements presented were taken.

[5] It should be stressed that these variations can be corrected by calibration using the position information given by the time difference between both strip ends (Figure 8).

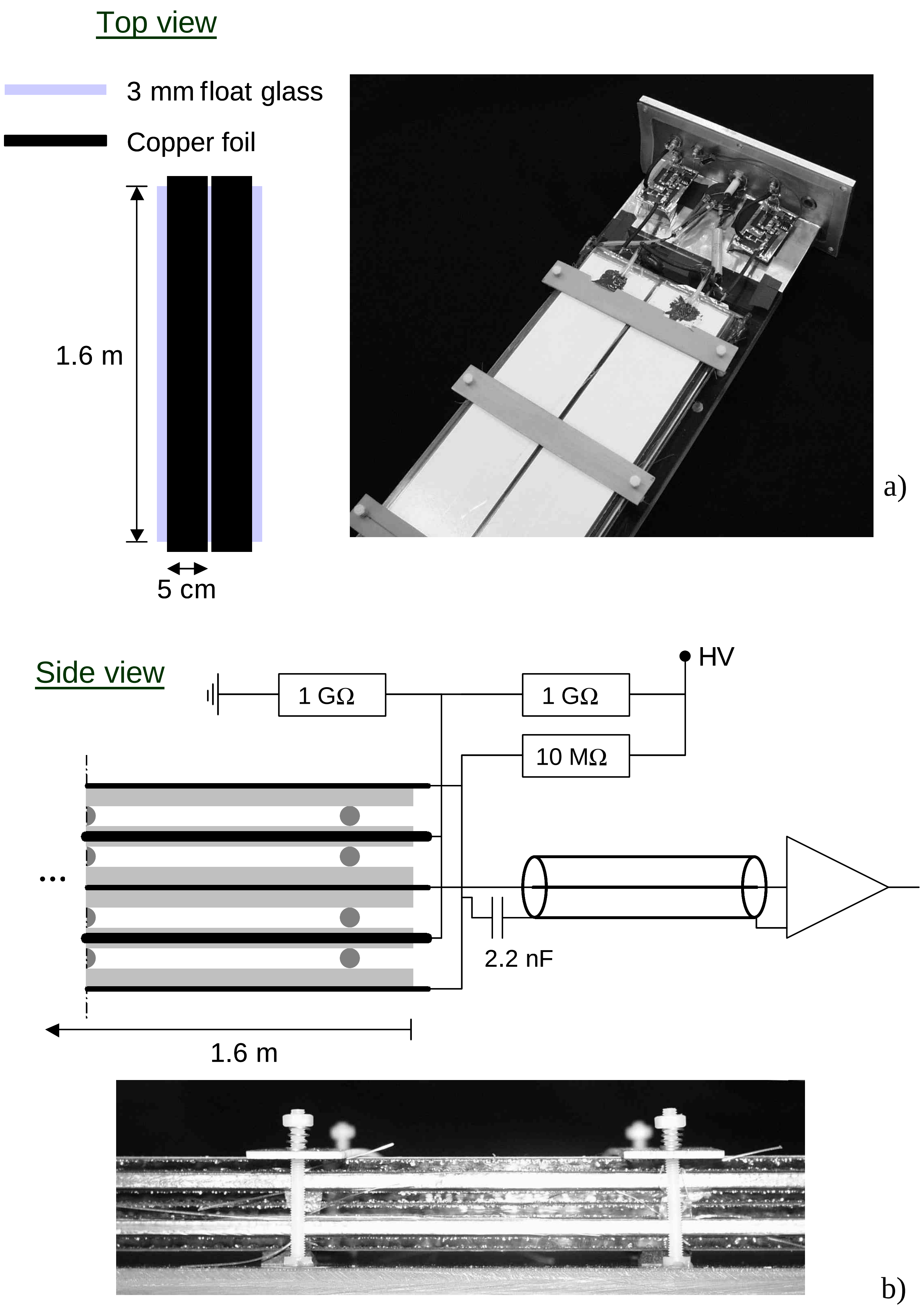

Figure 1: Pictures and schematic drawings of the detector.

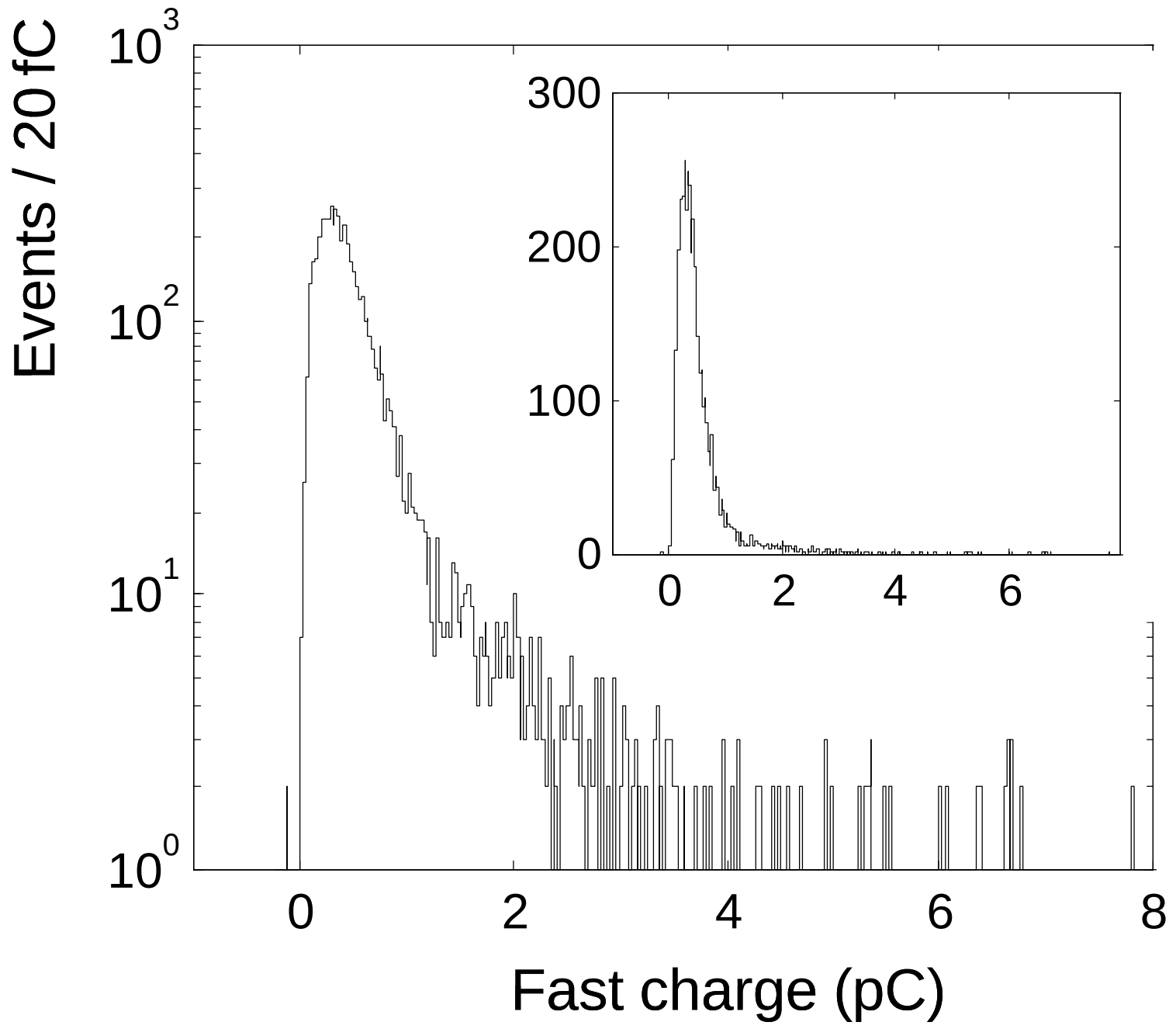

Figure 2: Typical fast charge distribution in logarithmic and linear (inset) scales.

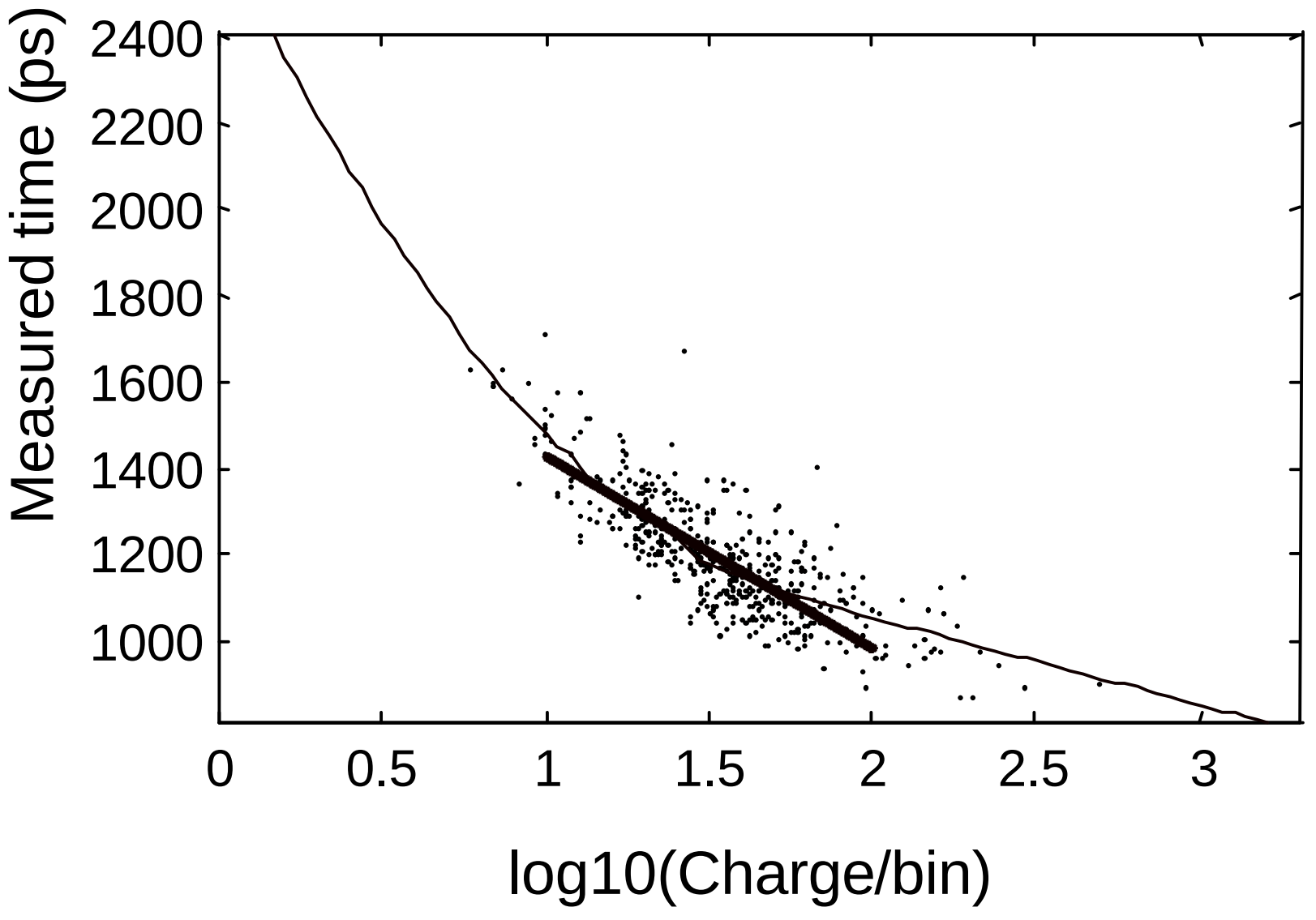

Figure 3: Typical time vs. log(charge) correlation plot, showing the calculated time-charge correction curve (thin line) and the average slope (thick line).

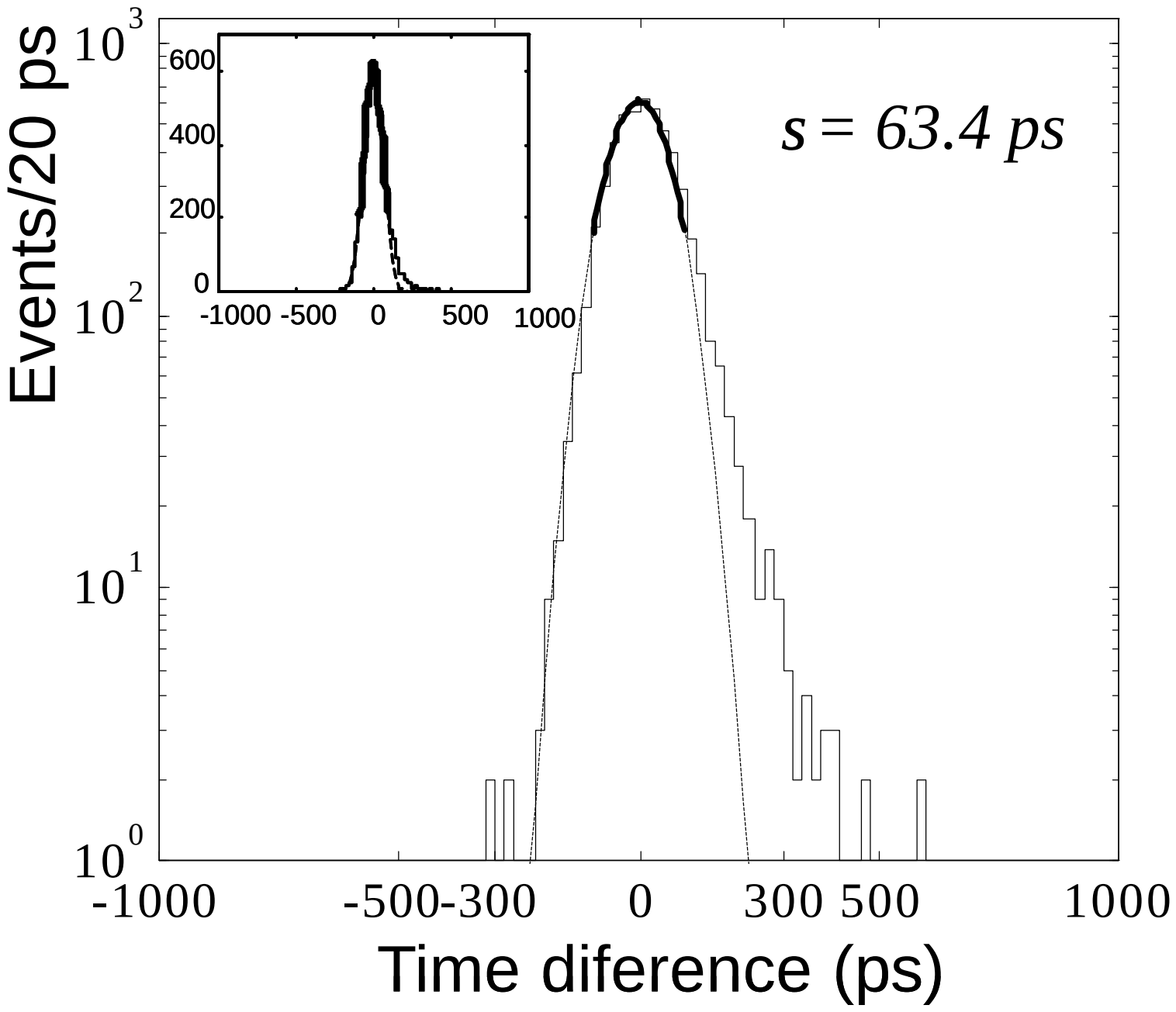

Figure 4: Typical time distribution (from strip B at X=-30 cm) after charge correction in logarithmic and linear (inset) scales. The thick line corresponds to a gaussian curve fitted within ±1.5 σ to determine the main resolution figure (after correction for the contribution of the start counters – see section 4.5 – the timing resolution for this example is σt=53 ps). The dashed line corresponds to the extension of the fitted gaussian to ±3.5 σ.

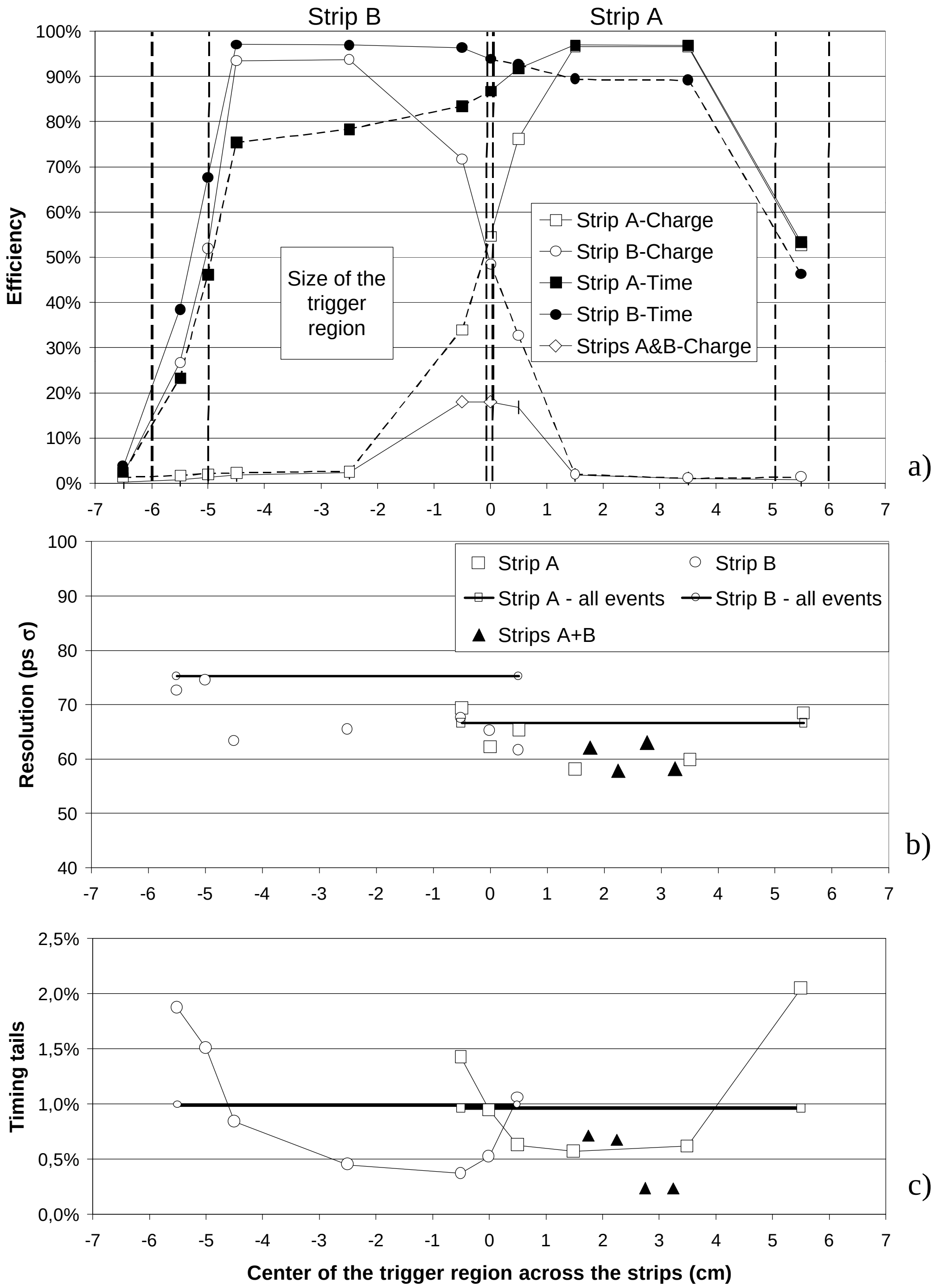

Figure 5: Several quantities of interest as a function of the position of the centre of the trigger region across the strips. a) Charge and time efficiency. The lower curve corresponds to the fraction of events that show a measurable amount of charge in both strips. The superimposed dashed lines indicate the position of the copper strips and the outer dotted line the edge of the glass plates. b) Timing resolution with separate analysis in each position and when all events for each strip are analysed simultaneously. The solid triangles correspond to data taken with both strips connected together. c) Same as b), for the timing tails.

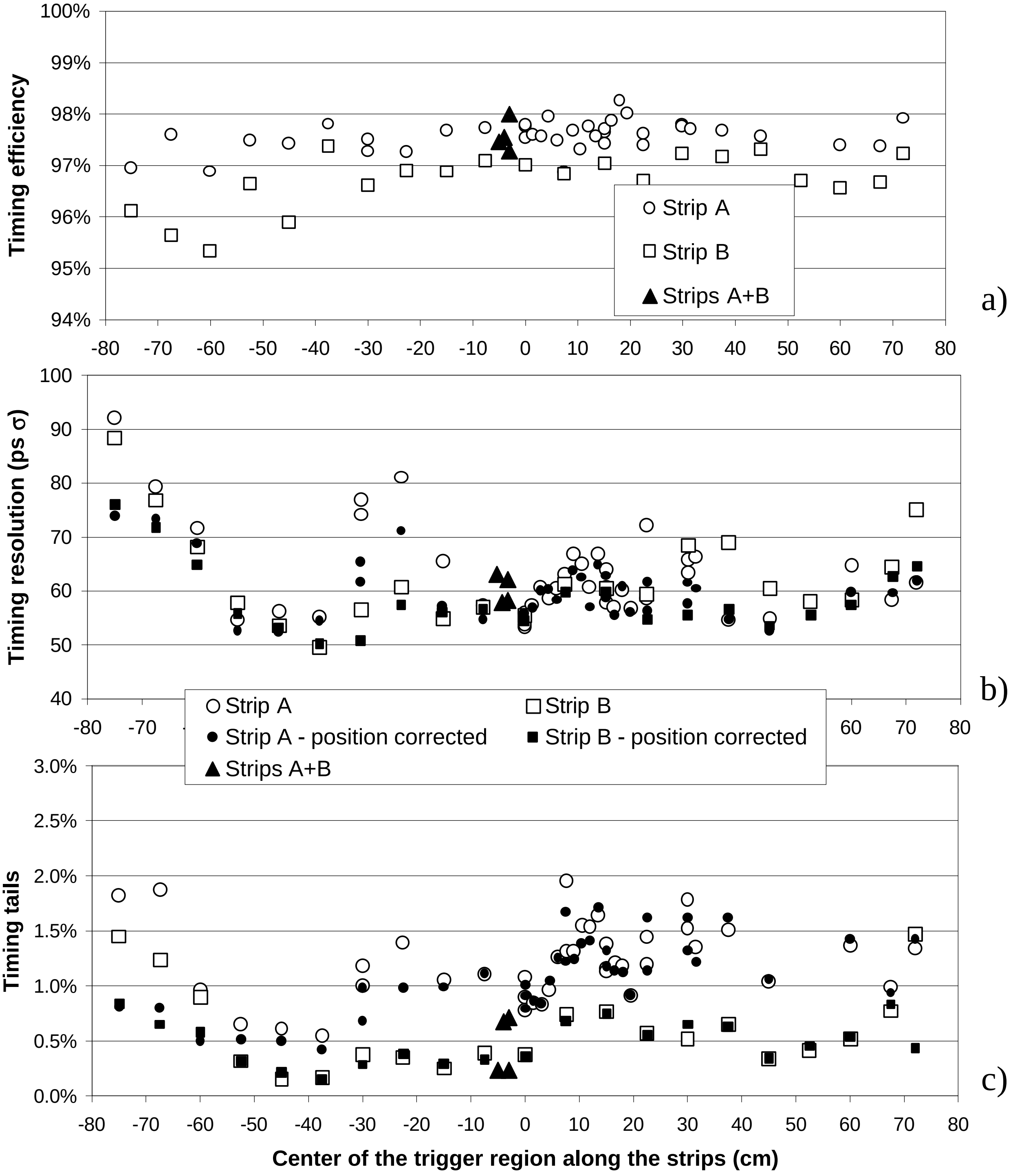

Figure 6: Several quantities of interest as a function of the position of the centre of the trigger region along the strips. a) Time efficiency, better than 95%; b) Time resolution with and without position correction, ranging from 50 to 90 ps σ (50 to 75 ps σ with position-dependent time correction). c) Timing tails with and without position correction, smaller than 2%. In all figures the solid triangles correspond to data taken with both strips connected together.

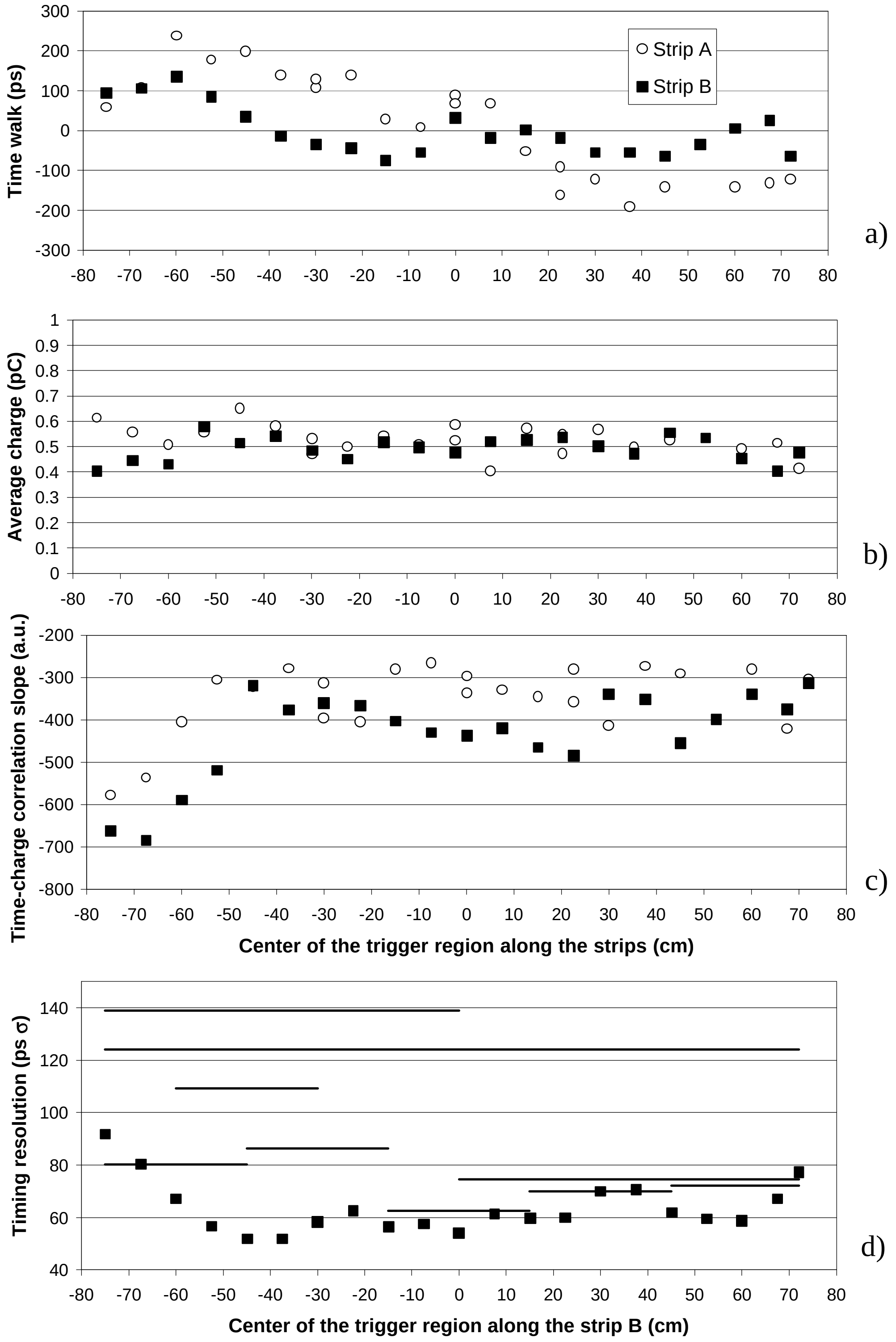

Figure 7: Several quantities related with the time-charge correlation curve (see Figure 3) are plotted as a function of the position of the trigger region along the strips. a) Variations of the average measured time, covering a range of 400 ps. b) Average measured charge, which shows little variation along the counter. c) The average slope shows considerable variations along the counter, particularly in the left hand side, where a poorer time resolution is also visible (Figure 6). d) Joint analysis of a data set containing an equal number events from each of the positions indicated by the extent of the horizontal lines: the right hand side is only weakly affected by position dependent effects, while the left hand side would require a more finely segmented calibration procedure.

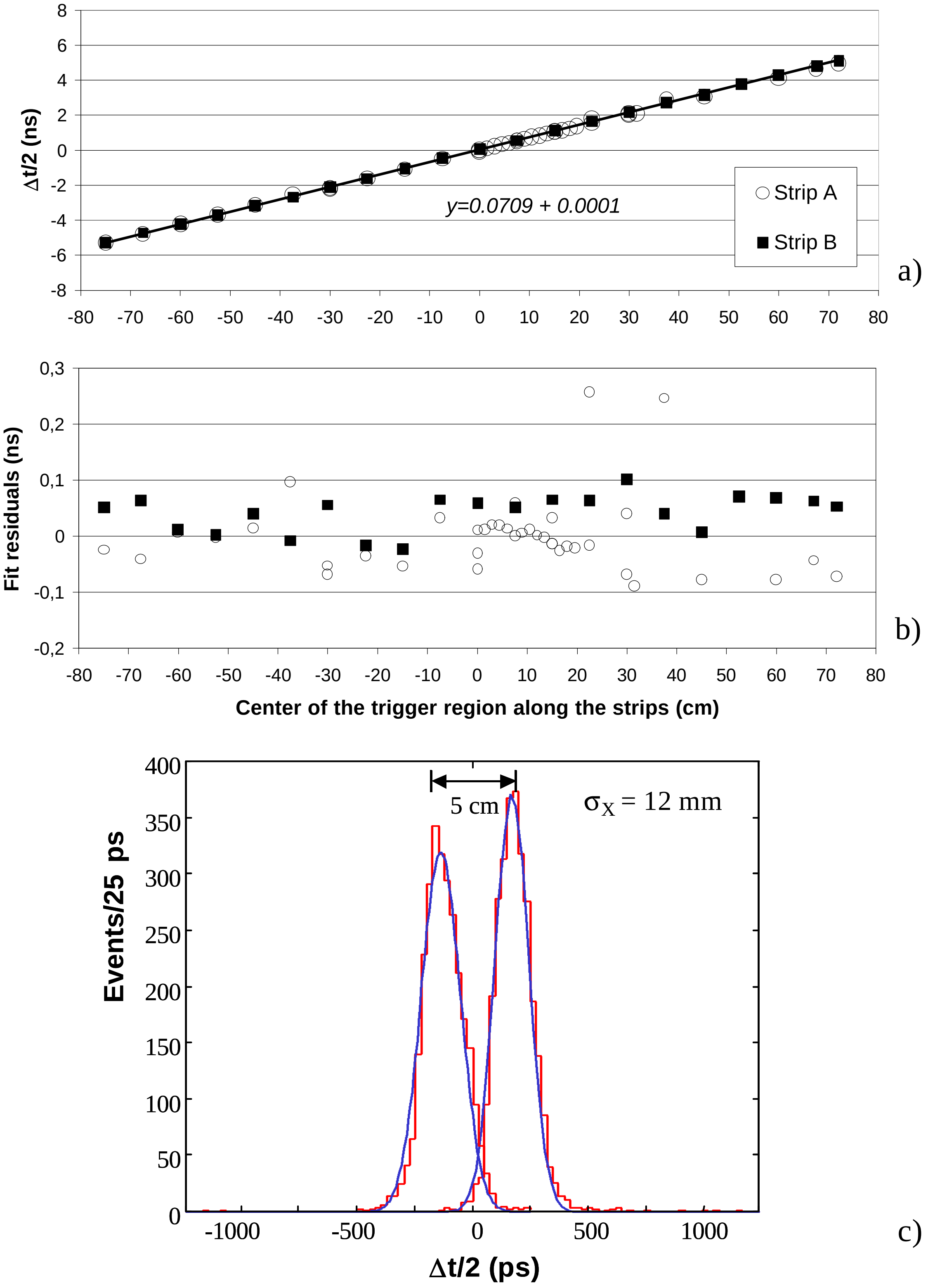

Figure 8: a) Time difference between both strip ends as a function of the position of the trigger region along the strips. There is an accurately linear dependency (evidenced by the small residues shown in b)), with a slope of 70.9 ps/cm, corresponding to a signal propagation velocity of 14.1 cm/ns. c) The width of the trigger region was reduced to 3 mm in the X direction and the spread of the time difference was compared with a measured displacement of 5.0 cm, yielding a position accuracy of 12 mm $\sigma$.

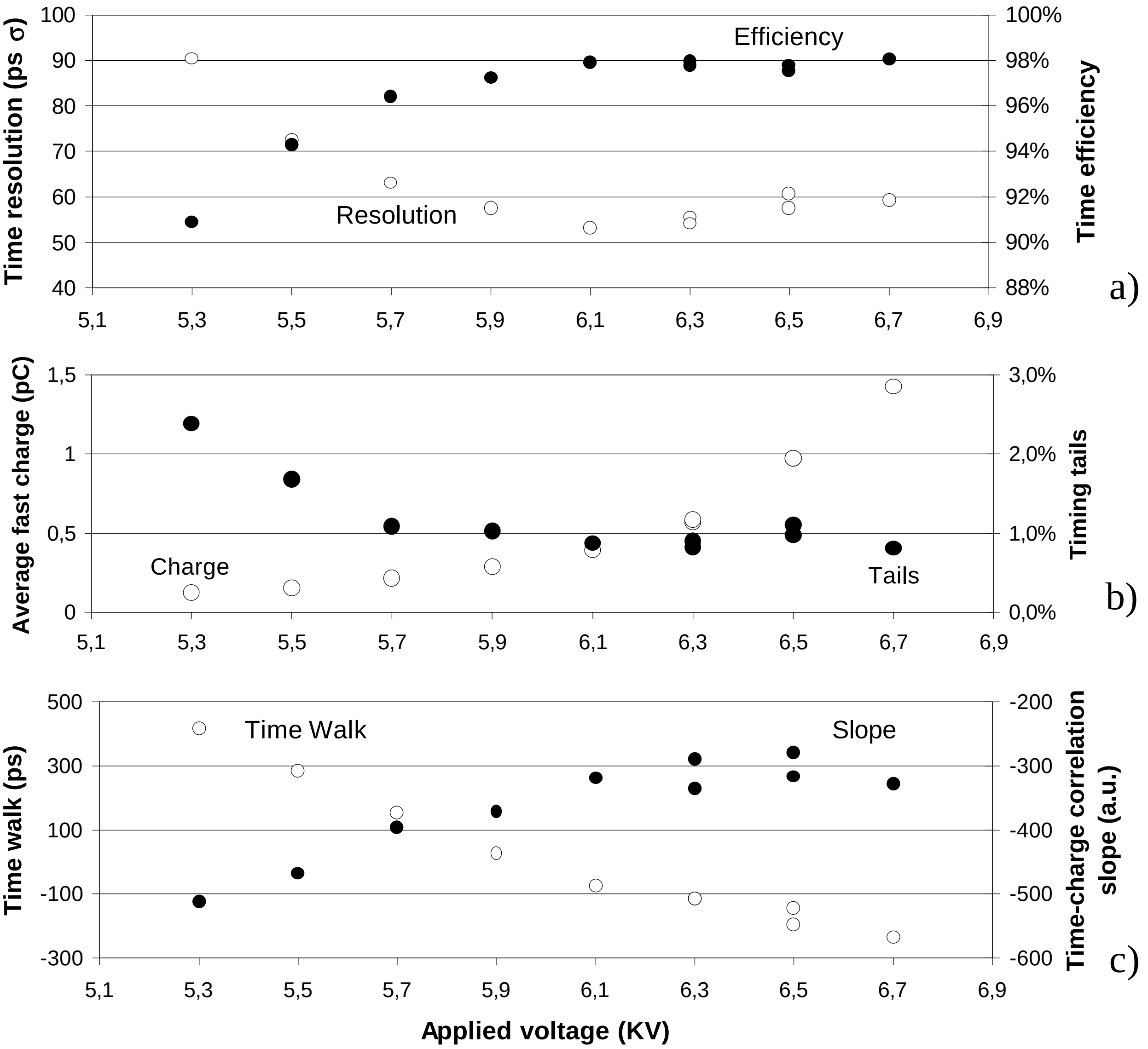

Figure 9: Several quantities of interest plotted as a function of the applied voltage. a) Time resolution and efficiency. b) Average amount of fast charge and the amount of timing tails. c) The variation of the absolute value of the measured time and the slope of the time vs. log(charge) correlation curve.

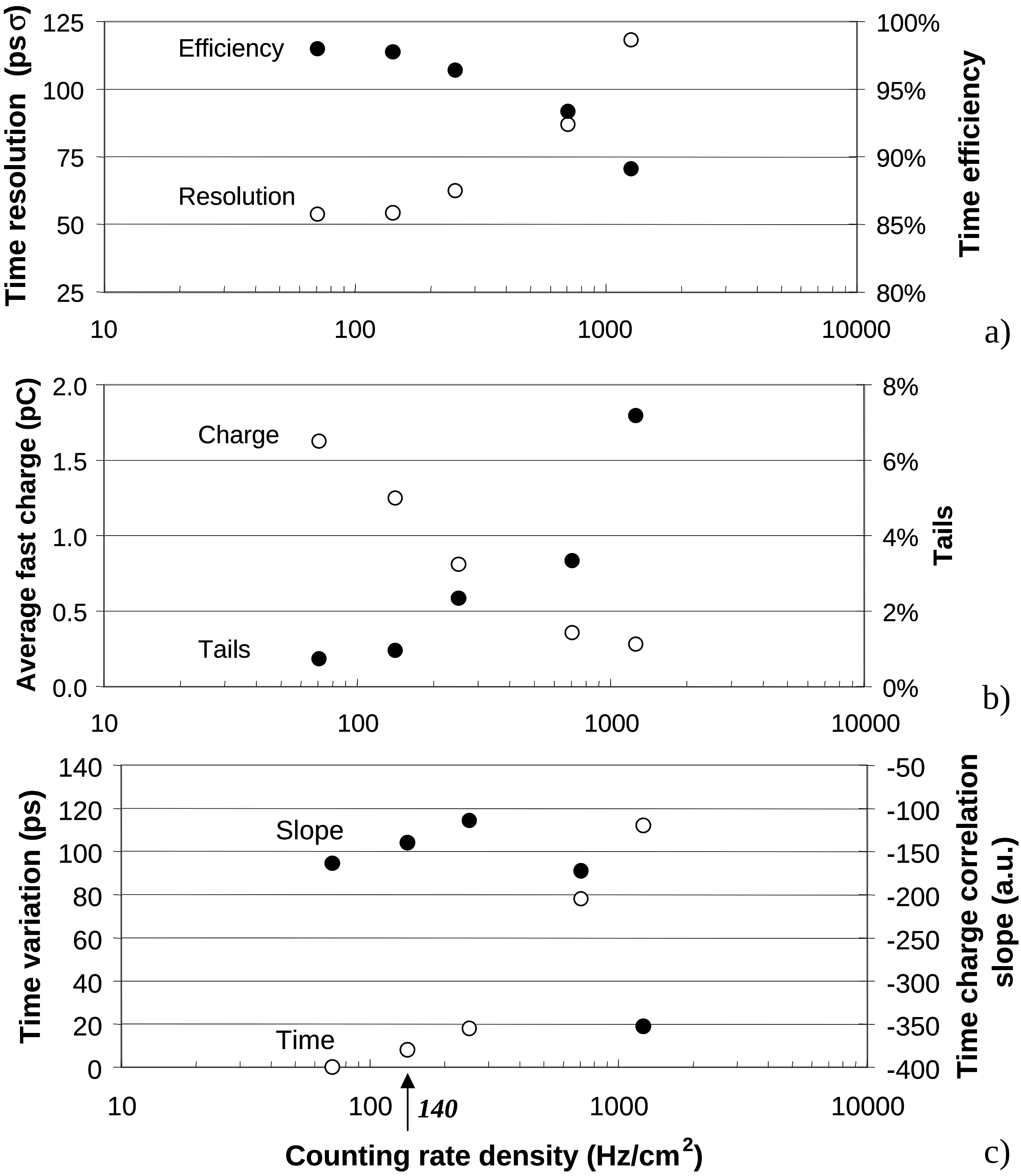

Figure 10: Several quantities of interest plotted as a function of the counting rate density. a) Time resolution and efficiency. b) Average fast charge and amount of timing tails. c) The variation of the absolute value of the measured time and the slope of the time vs. log(charge) correlation curve. The vertical arrow indicates the counting rate at which most of the measurements presented were taken.